\documentclass[english]{interact}
\usepackage[T1]{fontenc}
\usepackage[latin9]{inputenc}
\usepackage{refstyle}
\usepackage{mathrsfs}
\usepackage{amsmath}
\PassOptionsToPackage{version=3}{mhchem}
\usepackage{mhchem}

\makeatletter

\AtBeginDocument{\providecommand\figref[1]{\ref{fig:#1}}}
\RS@ifundefined{subsecref}
  {\newref{subsec}{name = \RSsectxt}}
  {}
\RS@ifundefined{thmref}
  {\def\RSthmtxt{theorem~}\newref{thm}{name = \RSthmtxt}}
  {}
\RS@ifundefined{lemref}
  {\def\RSlemtxt{lemma~}\newref{lem}{name = \RSlemtxt}}
  {}

\usepackage{pgfplots}
\usepackage{mathtools}
\usepackage[symbol]{footmisc}
\usepackage[sort&compress,numbers,merge]{natbib}
\bibpunct[, ]{[}{]}{,}{n}{,}{,}

\renewcommand{\figref}{\Figref}
\pgfplotsset{compat=1.3}
\usetikzlibrary{patterns}
\usetikzlibrary{plotmarks}

\pgfmathdeclarefunction{gauss}{2}{%
  \pgfmathparse{1/(#2*sqrt(2*pi))*exp(-((x-#1)^2)/(2*#2^2))}%
}

\makeatother

\usepackage{babel}
\begin{document}

\title{EOM-CC Methods with Approximate Triple Excitations for NEXAFS and
XPS}

\author{\name{Devin A. Matthews$^\ast$\thanks{$^\ast$E-mail address: damatthews@smu.edu}}\affil{Southern Methodist University, Dallas, TX 75275}}
\maketitle
\begin{abstract}
A number of iterative and perturbative approximations to the full
equation-of-motion coupled cluster method with single, double, and
triple excitations (EOM-CCSDT) are evaluated in the context of calculating
the K-edge core-excitation and core-ionization energies of several
small molecules. Several of these methods are found to accurately
reproduce the full EOM-CCSDT energies well, in particular the EOM-CCSD{*}
method which, when using the core-valence separation (CVS) ansatz,
scales rigorously with the sixth power of molecular size. The EOM-CCSDR(3)
method, which has been used previously to include triples effects
in a cost-effective manner was found to perform rather poorly, although
the precise cause has not been determined. These results highlight
that very accurate NEXAFS and XPS spectra for molecules with first-row
atoms can be computed at a cost not much larger than that for EOM-CCSD.

\end{abstract}

\section{Introduction}

The use of X-rays to investigate the structure and dynamics of molecules
and materials is long established \cite{hollanderXrayPhotoelectronSpectroscopy1970,fadleyAngleresolvedXrayPhotoelectron1984,koningsbergerXrayAbsorptionPrinciples1988,nordgrenSoftRayEmission1989},
but recent advances in X-ray sources, particular free-electron lasers,
have initiated a renaissance in X-ray absorption (NEXAFS, XPS) and
emission/scattering (XES/RIXS) applications \cite{fadleyXrayPhotoelectronSpectroscopy2010,milneRecentExperimentalTheoretical2014,bokhovenXRayAbsorptionXRay2016,bergmannXRayFreeElectron2017,krausUltrafastXraySpectroscopic2018}.
Computationally, a number of techniques are available to simulate
X-ray spectra with varying levels of fidelity and computational cost
\cite{normanSimulatingXraySpectroscopies2018,michelitschEfficientSimulationNearedge2019},
including DFT-based approaches such as time-dependent DFT (TD-DFT)
\cite{stenerTimeDependentDensity2003,tuSelfinteractioncorrectedTimedependentDensityfunctionaltheory2007,a.besleyTimedependentDensityFunctional2010,liangEnergySpecificLinearResponse2011,lestrangeCalibrationEnergySpecificTDDFT2015},
$\Delta$DFT ($\Delta$KS) \cite{stenerDensityFunctionalCalculations1995,takahashiFunctionalDependenceCoreexcitation2004,besleySelfconsistentfieldCalculationsCore2009,evangelistaOrthogonalityConstrainedDensity2013},
and transition-potential DFT (TP-DFT) \cite{stenerDensityFunctionalCalculations1995,huDensityFunctionalComputations1996,trigueroCalculationsNearedgeXrayabsorption1998},
as well as algebraic diagrammatic construction (ADC) methods \cite{barthTheoreticalCorelevelExcitation1985,trofimov2000core}.
More recently, there has been an increased interest in applying advanced
wavefunction-based electronic structure methods, in particular equation-of-motion
coupled cluster (EOM-CC) theory \cite{sekinoLinearResponseCoupledcluster1984,geertsenEquationofmotionCoupledclusterMethod1989,comeauEquationofmotionCoupledclusterMethod1993,stantonEquationMotionCoupledcluster1993}
to the X-ray regime \cite{nooijenDescriptionCoreExcitation1995,corianiCoupledclusterResponseTheory2012,corianiAsymmetricLanczosChainDrivenImplementationElectronic2012,kauczorCommunicationReducedspaceAlgorithm2013,pengEnergySpecificEquationofMotionCoupledCluster2015,corianiCommunicationXrayAbsorption2015,wolfProbingUltrafastPp2017,nascimentoSimulationNearEdgeXray2017,vidalNewEfficientEquationofMotion2019,zhengPerformanceDeltaCoupledClusterMethods2019,liuBenchmarkCalculationsKEdge2019,734c7d7e72234a548c61e39e075d7bd2,tsuruTimeresolvedNearedgeXray2019,fratiCoupledClusterStudy2019,faberResonantInelasticXray2019}.

Such applications are not without difficulties, however. Importantly,
the embedding of X-ray absorption resonances deep in the valence continuum
leads to severe convergence issues; initial studies utilizing energy-windowing
\cite{pengEnergySpecificEquationofMotionCoupledCluster2015}, damped
response \cite{corianiCoupledclusterResponseTheory2012,kauczorCommunicationReducedspaceAlgorithm2013},
or Lanczos-based \cite{corianiAsymmetricLanczosChainDrivenImplementationElectronic2012}
solvers have given way to the core-valence separation \cite{cederbaumManybodyTheoryCore1980},
introduced to EOM-CC by Coriani and Koch \cite{corianiCommunicationXrayAbsorption2015}.
This approximation has proven highly resilient and accurate, and has
enabled routine use of EOM-CC in X-ray calculations. Just as important,
though, is the issue of orbital relaxation. As the presence of the
core hole in the excited (or ionized) state leads to severe contraction
of the valence orbitals, the ground state orbitals prove to be a poor
reference. EOM-CC corrects for orbital relaxation via electron correlation,
but three-electron correlation (e.g. the EOM-CCSDT method \cite{hirataHighorderDeterminantalEquationofmotion2000})
is necessary to fully account for such a large relaxation effect \cite{liuBenchmarkCalculationsKEdge2019,734c7d7e72234a548c61e39e075d7bd2}.
The full EOM-CCSDT method is much too expensive for routine calculations
or indeed for any calculation of molecules beyond a handful of atoms
due to its steep scaling with molecular size. Instead, one may opt
to include the effect of triple excitations in an approximate fashion,
using either an iterative \cite{christiansenResponseFunctionsCC31995,wattsEconomicalTripleExcitation1995,wattsIterativeNoniterativeTriple1996}
or a perturbative \cite{wattsEconomicalTripleExcitation1995,christiansenPerturbativeTripleExcitation1996,stantonSimpleCorrectionFinal1996,saehApplicationEquationofmotionCoupled1999,matthewsNewApproachApproximate2016}
model.

At present, calculations beyond the CCSD level are rather sparsely
represented in the literature; perhaps the best example is the use
of the EOM-CCSDR(3) method \cite{christiansenPerturbativeTripleExcitation1996}
in the work of Coriani et al \cite{corianiAsymmetricLanczosChainDrivenImplementationElectronic2012,corianiCoupledclusterResponseTheory2012}.
In this article, we show that several computationally efficient approximations
to full EOM-CCSDT are effective in recovering nearly all of the residual
orbital relaxation energy (i.e. the relaxation not incorporated in
EOM-CCSD).

\section{Theory}

\subsection{Core-Valence Separated Equation-of-Motion Coupled Cluster}

The equation-of-motion coupled cluster (EOM-CC) approach \cite{sekinoLinearResponseCoupledcluster1984,geertsenEquationofmotionCoupledclusterMethod1989,comeauEquationofmotionCoupledclusterMethod1993,stantonEquationMotionCoupledcluster1993}
computes the excitation energy $\omega_{EE}$ as the eigenvalue of
a transformed Hamiltonian (EOMEE-CC),
\begin{align*}
\omega_{EE}\hat{R}_{EE}|\Phi_{0}\rangle & =[\bar{H},\hat{R}_{EE}]|\Phi_{0}\rangle=\bar{H}_{o}\hat{R}_{EE}|\Phi_{0}\rangle\\
\hat{R}_{EE} & =\hat{R}_{0}+\hat{R}_{1}+\hat{R}_{2}+\cdots\\
 & =r_{0}+\sum_{ai}r_{i}^{a}a_{a}^{\dagger}a_{i}+\frac{1}{4}\sum_{abij}r_{ij}^{ab}a_{a}^{\dagger}a_{b}^{\dagger}a_{j}a_{i}+\cdots
\end{align*}
where $|\Phi_{0}\rangle$ is the reference determinant (usually Hartree-Fock)
and $\bar{H}=e^{-\hat{T}}\hat{H}e^{\hat{T}}$ is the coupled cluster
transformed Hamiltonian. The ``open'' transformed Hamiltonian $\bar{H}_{o}=\bar{H}-\langle0|\bar{H}|0\rangle$.
Following the usual convention, occupied orbitals are denoted by $ijk\ldots$
while virtual orbitals are denoted by $abc.\ldots$ The $\hat{T}$
operator is the well-known cluster operator which is determined by
the ground state coupled cluster equations,
\begin{align*}
\langle\Phi_{ij\ldots}^{ab\ldots}|\bar{H}_{o}|\Phi_{0}\rangle & =0\\
\hat{T} & =\hat{T}_{1}+\hat{T}_{2}+\cdots\\
 & =\sum_{ai}t_{i}^{a}a_{a}^{\dagger}a_{i}+\frac{1}{4}\sum_{abij}t_{ij}^{ab}a_{a}^{\dagger}a_{b}^{\dagger}a_{j}a_{i}+\cdots
\end{align*}
where $\langle\Phi_{ij\ldots}^{ab\ldots}|$ is an excited determinant.
EOM-CC is also applicable to ionization energies $\omega_{IP}$ (EOMIP-CC)
through the use of a non-number conserving operator,
\begin{align*}
\omega_{IP}\hat{R}_{IP}|\Phi_{0}\rangle & =\bar{H}_{o}\hat{R}_{IP}|\Phi_{0}\rangle\\
\hat{R}_{IP} & =\hat{R}_{1}+\hat{R}_{2}+\cdots\\
 & =\sum_{i}r_{i}a_{i}+\frac{1}{2}\sum_{aij}r_{ij}^{a}a_{a}^{\dagger}a_{j}a_{i}+\cdots
\end{align*}
The truncation of the $\hat{T}$ operator along with an equivalent
truncation of the $\hat{R}$ ($EE$ or $IP$) operators defines a
particular canonical EOM-CC method. For example, the truncation $\hat{T}=\hat{T}_{1}+\hat{T}_{2}$
and $\hat{R}=\hat{R}_{1}+\hat{R}_{2}$ gives EOM-CCSD (note that $\hat{R}_{0}$
does not enter the excitation energy equations due to the fact that
the coupled cluster equations have been solved).

The core-valence separation (CVS) \cite{cederbaumManybodyTheoryCore1980,corianiCommunicationXrayAbsorption2015}
is a technique that decouples the core excitation or ionization spectrum
from the valence continuum. As core states are resonances embedded
in the valence continuum, the core-valence separation allows for direct
calculations as if the core states were bound and avoid complications
from convergence problems and unphysical couplings \cite{liuBenchmarkCalculationsKEdge2019}.
The CVS has been shown to introduce negligible error on the computed
excitation and ionization energies for first-row atoms \cite{corianiCommunicationXrayAbsorption2015,liuBenchmarkCalculationsKEdge2019}.
In the context of EOM-CC, the CVS is equivalent to the restriction
that only operators in $\hat{R}_{EE}$ and $\hat{R}_{IP}$ that contain
at least one core orbital index are retained. In the present calculations,
we also require that this core orbital index corresponds to only the
selected atomic number. For example, a CVS-EOM-CC calculation of the
oxygen K-edge in CO would neglect operators with only valence or C
$1s$ occupied indices. This requirement is added to separate higher
K edges from the core ionization continuum of a lower-energy edge.
In larger molecules, it would also be possible to require that only
core orbitals of one or a small number of symmetry-unique atoms are
included in the core-valence separation. Since such core orbitals
are highly localized, this is expected to be a very good approximation,
and in theory dramatically reduces the number of excitation or ionization
operators needed.

\subsection{Methods with Approximate Triple Excitations}

The EOMEE-CCSDT method \cite{hirataHighorderDeterminantalEquationofmotion2000}
canonically scales as $\mathscr{O}(n^{8})$ in both the ground state
and excited state calculations. EOMIP-CCSDT reduces the scaling for
the excited state to $\ensuremath{\mathscr{O}}(n^{7})$ but has the
same ground state cost. For applications on realistic molecules, then,
an approximation to the triples excitations that reduces the scaling
down to $\mathscr{O}(n^{7})$ or better yet to $\mathscr{O}(n^{6}$)
is highly desirable. As triples effects are important for reaching
sub-eV accuracy for core-excited and core-ionized states (see \cite{liuBenchmarkCalculationsKEdge2019,734c7d7e72234a548c61e39e075d7bd2}
and the results in this work), such an approximation is critical for
accurate calculations. The approximations assessed here span two different
approaches: iterative approximations and non-iterative (perturbative)
approximations.

The iterative approximations share the common feature that they neglect
one or more terms in the transformed Hamiltonian $\bar{H}_{o}$. For
example, the EOM-CCSDT-3 method \cite{wattsIterativeNoniterativeTriple1996}
includes all terms except those that scale as $\mathscr{O}(n^{8})$,
or equivalently the terms of the form $\langle T|\tilde{W}|T\rangle$
where $|T\rangle$ represents the triple excitation manifold and $\tilde{W}$
is the two-particle part of $\bar{H}_{o}$. Importantly, these methods
apply the same approximation of the transformed Hamiltonian in the
ground state and excited state calculations. This ensures that the
EOM-CC excitation energies are equivalent to the linear response coupled
cluster (LR-CC) poles of the ground state wavefunction. The iterative
approximations used here are EOM-CCSDT-1 \cite{wattsEconomicalTripleExcitation1995},
-CCSDT-3, and -CC3 \cite{christiansenResponseFunctionsCC31995}. The
related EOM-CCSDT-1b and -CCSDT-2 methods are not expected to provide
any additional insights into the behavior of such methods for X-ray
spectroscopy.

Perturbative approximations approach the problem from a different
direction: instead of approximating the transformed Hamiltonian, they
approximate the energy directly, and retain only important terms as
determined by some partitioning of the Hamiltonian. The CCSD(T) method
is the most well-known perturbative method for the ground state, but
unlike the iterative approximations, CCSD(T) provides no approximate
wavefunction or transformed Hamiltonian which can be used in the LR-CC
or EOM-CC equations. The EOM-CCSD{*} method \cite{stantonSimpleCorrectionFinal1996,saehApplicationEquationofmotionCoupled1999}
was developed in analogy to CCSD(T), but working from the final excited
state energy in the perturbation expansion. In order to balance the
treatment of triple excitations in the ground and excited states,
the EOM-CCSDR(3) \cite{christiansenPerturbativeTripleExcitation1996}
and EOM-CCSD(T)(a) \cite{matthewsNewApproachApproximate2016} methods
were introduced. The latter explicitly constructs an approximate transformed
Hamiltonian which is used in the excited state calculation, in close
analogy to the iterative approximations. However, this requires an
iterative calculation of the triples in the excited state, and so
the EOM-CCSD(T)(a){*} method \cite{matthewsNewApproachApproximate2016}
was also introduced which blends the EOM-CCSD(T)(a) and EOM-CCSD{*}
approaches.

\subsection{Scaling Reduction}

Almost all of the approximate approaches scale as $\mathscr{O}(n^{7})$
for the ground state, with the exception of EOM-CCSD{*}. Of course,
the leading-order cost is a one-time correction for EOM-CCSDR(3) and
EOM-CCSD(T)(a) (``star'' and ``no star''), while for the iterative
methods this cost must be payed each iteration of the coupled cluster
equations. In the excited state, these methods also scale as $\mathscr{O}(n^{7})$
for valence excitations. Within the CVS, though, we can theoretically
restrict the $\hat{R}$ operators to only those allowable (with at
least one target core orbital) and reduce the scaling by one order
of $n$. In the EOM-CCSD part of the calculation, this scaling reduction
has only been recently achieved by Vidal et al. \cite{vidalNewEfficientEquationofMotion2019}
and is not realized in the implementation in this work. However, the
restriction of the triples excitations has been implemented. This
is made possible by the non-orthogonal spin-adaptation approach \cite{matthewsNonorthogonalSpinadaptationCoupled2015}
used in CFOUR \cite{stantonCFOURCoupledClusterTechniques}, where
the triples (and quadruples) amplitudes are stored as a set of ``virtual
blocks''. For example, the $\hat{R}_{3}$ amplitudes are stored as
a set of dense tensors $\left\{ r_{abc}\right\} _{ijk}$, one for
each symmetry-unique combination of the occupied orbital indices.
Because the occupied indices are already treated separately from the
virtuals and take advantage of sparsity due to permutational symmetry,
this approach was naturally extensible to the CVS by simply filtering
out all $ijk$ indices that do not contain at least one target core
orbital. In this way, the scaling of all of the approximate approaches
except for EOM-CCSD(T)(a) is reduced to $\mathscr{O}(n^{6})$ for
the excited state calculation, and even for EOM-CCSD(T)(a) the calculation
of ionization energies scales as at most $\mathscr{O}(n^{6})$. The
methods used here should then be suitable for molecular sizes in the
range that is currently accessible via CCSD(T), CC3, etc., with EOM-CCSD{*}
potentially scalable to even larger systems due to strict $\mathscr{O}(n^{6})$
scaling.\footnote{Of course, all of this analysis assumes treatment of core-excited
and core-ionized states within the CVS approximation and with only
a small number of core orbitals targeted. In the valence case, for
example, scaling is not reduced.}

\section{Computational Approach}

In the following benchmarks, EOM-CCSDT is used as the target method,
as previous work has shown it to be accurate in comparison to experiment
to within about 0.1 eV \cite{liuBenchmarkCalculationsKEdge2019,734c7d7e72234a548c61e39e075d7bd2}
with the aCVQZ (aug-cc-pCVQZ \cite{dunningGaussianBasisSets1989a,kendallElectronAffinitiesFirst1992,woonGaussianBasisSets1995})
basis. A limited number of calculations with EOM-CCSDTQ have also
shown only small (<0.1 eV) residual correlation effects. Calculations
at the EOM-CCSDT level were carried out using a development version
of the CFOUR program package \cite{stantonCFOURCoupledClusterTechniques}
on the core ionization energies and the first four (vertical) core
excitation energies for each K edge of $\ce{H2O}$, CO, $\ce{NH3}$,
HCN, and $\ce{C2H4}$, using the aCVTZ basis set except for $\ce{H2O}$
which used the aCVQZ basis set. The computational geometries are listed
in the Supplemental Information. For ethylene, the average of the
nearly-degenerate gerade and ungerade transitions was used. EOM-CCSD,
--CCSD{*}, --CCSDR(3), --CCSD(T)(a){*}, --CCSDT-1, --CCSDT-3,
and --CC3 calculations with the same basis sets were then performed
on this test set. Additionally, EOM-CCSD(T)(a) (with an iterative
triples correction for the excited state) was used to calculate all
core ionization energies. All EOM calculations utilized the CVS with
only a single core orbital targeted, except for $\ce{C2H4}$ where
molecular symmetry requires that both C $1s$ orbitals are included
in the target space. We also performed direct two-determinant coupled
cluster singles and doubles (TD-CCSD \cite{kucharskiHilbertSpaceMultireference1991,balkovaCoupledclusterMethodOpenshell1992})
calculations on all core-excited states, as all of these states are
well-described as two-determinant open-shell singlets. Most states
were accessible at this level of theory, however the 4s and 5s Rydberg
states of HCN, CO, and $\ce{NH3}$ could not be converged with TD-CCSD.
We have also excluded $\ce{C2H4}$ from the TD-CCSD treatment as a
broken-symmetry localized core-hole SCF solution is required to fully
relax the orbitals in this case \cite{zhengPerformanceDeltaCoupledClusterMethods2019}.
The difference between the TD-CCSD energies and the CCSD ground state
energy is denoted by $\Delta$CCSD, in analogy with $\Delta$SCF methods.
The $\Delta$CCSD ionization energies correspond to the difference
between an ROCCSD calculation of the core-ionized state and the CCSD
ground-state energy. Note that the TD-CCSD and ROCCSD calculations
required that the ``dangerous denominators'' be removed;\cite{nooijenDescriptionCoreExcitation1995,zhengPerformanceDeltaCoupledClusterMethods2019}
in the present implementation these denominators $\Delta_{ij}^{ab}$
are identified as those for which $\min(\varepsilon_{i},\varepsilon_{j})>\min(\varepsilon_{a},\varepsilon_{b})$.
No electrons are frozen in any of these calculations.

\section{Results and Discussion}

\subsection{Ionization Energies}

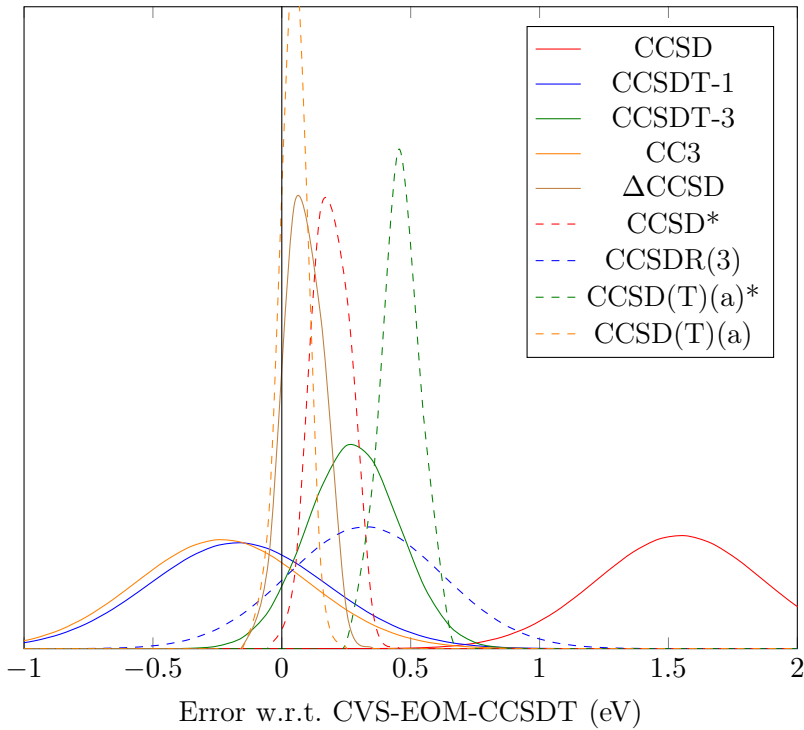
\begin{figure}
\begin{centering}
\thispagestyle{empty}
\begin{tikzpicture}
\begin{axis}[
scale=1.5,
samples=100,smooth,
xmin=-1, xmax=2,
ymin=0, ymax=7,
ytick=\empty,
xlabel={Error w.r.t. CVS-EOM-CCSDT (eV)},
legend style={legend pos=north east}
]
\addplot[draw=red] {gauss(1.544857143,0.323325324)};
\addlegendentry{CCSD}
\addplot[draw=blue] {gauss(-0.170857143,0.345274395)};
\addlegendentry{CCSDT-1}
\addplot[draw=black!50!green] {gauss(0.277142857,0.178180028)};
\addlegendentry{CCSDT-3}
\addplot[draw=orange] {gauss(-0.231428571,0.335402043)};
\addlegendentry{CC3}
\addplot[draw=brown] {gauss(0.084764248,0.073313144)};
\addlegendentry{$\Delta$CCSD}
\addplot[draw=red,dashed] {gauss(0.190428571,0.071397993)};
\addlegendentry{CCSD*}
\addplot[draw=blue,dashed] {gauss(0.323142857,0.299922099)};
\addlegendentry{CCSDR(3)}
\addplot[draw=black!50!green,dashed] {gauss(0.461142857,0.072915663)};
\addlegendentry{CCSD(T)(a)*}
\addplot[draw=orange,dashed] {gauss(0.044142857,0.051962702)};
\addlegendentry{CCSD(T)(a)}
\addplot[draw=black] coordinates {(0,-1) (0,100)};
\end{axis}
\end{tikzpicture}
\par\end{centering}
\caption{\label{fig:ip}Normal error distributions for core ionization energies.
Iterative methods are denoted by solid lines, and perturbative methods
by dashed lines. All methods except $\Delta$CCSD are core-valence
separated EOM, e.g. ``CCSD'' = CVS-EOM-CCSD.}
\end{figure}

The aggregate results are summarized as normal error distributions
in \figref{ip}, while the full results are available in the Supplementary
Information. In this format, the ionization energy error for method
$X$, $\omega_{IP}(X)-\omega_{IP}(\text{CVS-EOM-CCSDT})$, is modeled
by a normal distribution and plotted as a normalized Gaussian function.
In the following discussion, the ``short'' name for each method
will be used for brevity, e.g. CVS-EOM-CCSD will be denoted as simply
CCSD. The exception is the $\Delta$CCSD method which is not an equation-of-motion
approach, and should not be confused with ``CCSD''.

From the error distributions, we can see that the most accurate method
is CCSD(T)(a), without the perturbative ``star'' correction. This
is not unexpected as this method performs essentially full CCSDT for
the excited state, with only the ground state triple excitations being
approximated. Because the most important correlation effects for the
core hole (orbital relaxation and core hole correlation) occur only
in the excited state, this explains its superior performance. Approximation
of the ground state has much less of an effect as the majority of
the valence correlation error cancels between the excited and ground
states. On the other end of the spectrum, CCSD predicts core ionization
energies almost uniformly more than 1 eV above CCSDT. $\Delta$CCSD
significantly improves on CCSD as it takes orbital relaxation into
account explicitly rather than through correlation in the excited
state. In contrast, the iterative triples approximations, with the
exception of CCSDT-3, fail to improve the unsystematic error. The
95\% confidence interval (CI) for each of these methods is greater
than 1.2 eV. CCSDT-3 is an improvement, but only a marginal one as
the CI only drops to $\sim0.7$ eV. Finally, the perturbative approximations
CCSD{*} and CCSD(T)(a){*} both perform very well, with CIs below 0.3
eV, although the mean error is as large as 0.5 eV. Interestingly,
CCSDR(3) is not able to improve on the iterative methods. Because
full CCSD(T)(a) scales iteratively as $\mathscr{O}(n^{7})$ for core
excitation calculations, it is really only a viable candidate for
core ionization energies. However, its superior performance for absolute
ionization energies opens the possibility that it could be used as
an internal correction for systematic errors (systematic within the
X-ray spectrum of a specific molecular K edge--this is a more powerful
correction than a simple rigid shift). To this end, we also defined
the ``corrected'' CCSD{*} method for core-excited states,
\[
\omega_{EE}(\text{CCSD* corr.})=\omega_{EE}(\text{CCSD*})+\omega_{IP}(\text{CCSD(T)(a)})-\omega_{IP}(\text{CCSD*})
\]

\subsection{Vertical Excitation Energies}

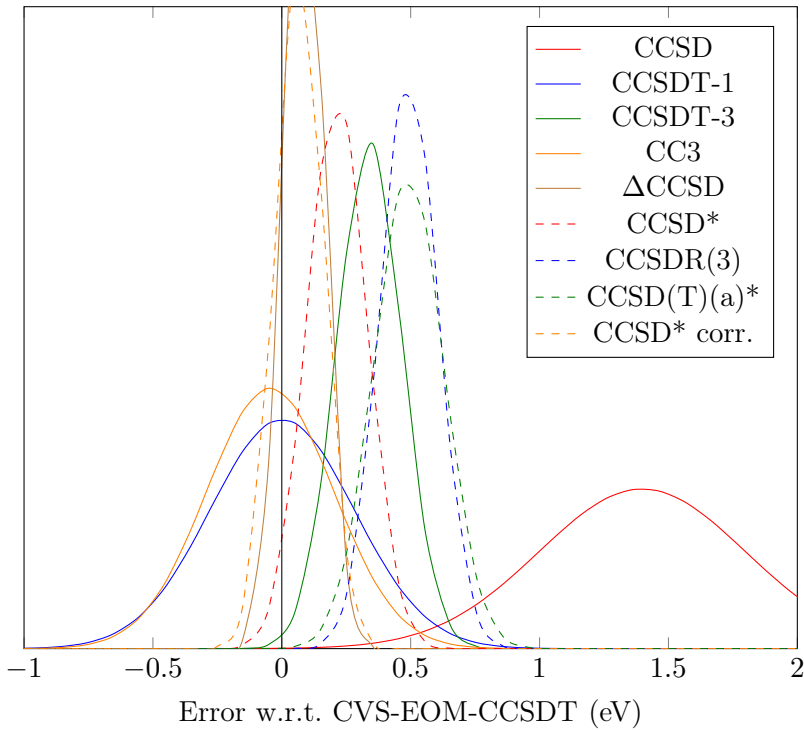
\begin{figure}
\begin{centering}
\thispagestyle{empty}
\begin{tikzpicture}
\begin{axis}[
scale=1.5,
samples=100,smooth,
xmin=-1, xmax=2,
ymin=0, ymax=4,
ytick=\empty,
xlabel={Error w.r.t. CVS-EOM-CCSDT (eV)},
legend style={legend pos=north east}
]
\addplot[draw=red] {gauss(1.398370133,0.401126705)};
\addlegendentry{CCSD}
\addplot[draw=blue] {gauss(0.00119655,0.279436861)};
\addlegendentry{CCSDT-1}
\addplot[draw=black!50!green] {gauss(0.340375,0.126187405)};
\addlegendentry{CCSDT-3}
\addplot[draw=orange] {gauss(-0.044946429,0.245882171)};
\addlegendentry{CC3}
\addplot[draw=brown] {gauss(0.083879934,0.078573783)};
\addlegendentry{$\Delta$CCSD}
\addplot[draw=red,dashed] {gauss(0.214061156,0.115800254)};
\addlegendentry{CCSD*}
\addplot[draw=blue,dashed] {gauss(0.492725425,0.111320507)};
\addlegendentry{CCSDR(3)}
\addplot[draw=black!50!green,dashed] {gauss(0.492683499,0.13499577)};
\addlegendentry{CCSD(T)(a)*}
\addplot[draw=orange,dashed] {gauss(0.067775442,0.092944912)};
\addlegendentry{CCSD* corr.}
\addplot[draw=black] coordinates {(0,-1) (0,100)};
\end{axis}
\end{tikzpicture}
\par\end{centering}
\caption{\label{fig:absolute}Normal error distributions for vertical core
excitation energies. Iterative methods are denoted by solid lines,
and perturbative methods by dashed lines. All methods except $\Delta$CCSD
are core-valence separated EOM, e.g. ``CCSD'' = CVS-EOM-CCSD.}
\end{figure}

Similar normal error distributions for all methods (now replacing
CCSD(T)(a) with CCSD{*} corr.) are given for vertical core excitation
energies in \figref{absolute}. The CCSD and $\Delta$CCSD methods
perform essentially identically to the ionization energy case (note
that while the vertical axis is arbitrary, it is slightly more compressed
in this figure). CCSDT-1 and CC3 also show minimal change, performing
similarly to CCSD in terms of standard distribution albeit with essentially
zero mean error. CCSD{*} and CCSD(T)(a){*} also continue to perform
quite well, with CCSDT-3 improving slightly so that it roughly matches
the perturbative approximations in accuracy. The corrected CCSD{*}
method also illustrates that the correction from the CCSD(T)(a) ionization
energy can remove some systematic errors from CCSD{*}, as the mean
error is shifted to almost zero. Additionally, the standard deviation
of the CCSD{*} error drops slightly, showing that this correction
is indeed leading to a more sophisticated error cancellation than
a rigid shift. Finally, the most significant change from the ionization
energy case is that the error in CCSDR(3) is dramatically improved
such that it is essentially indistinguishable from CCSD(T)(a){*}.

Of course, both the ionization and vertical excitation energies computed
here are not necessarily the proper observables in the context of
X-ray spectroscopy. In XPS, the chemical shift, i.e. the change in
core ionization energy from some reference molecule has been shown
to be less sensitive to errors in correlation energy. For CVS-EOM-CCSD,
Liu et al. \cite{liuBenchmarkCalculationsKEdge2019} found that the
mean absolute error decreased by approximative a factor of 5--6 when
considering chemical shifts rather than absolute ionization energies,
but that the standard deviation only decreased by about 20\%. Zheng
et al. \cite{zhengPerformanceDeltaCoupledClusterMethods2019} found
similar results for $\Delta$CCSD, while in both studies the addition
of triples effects gave essentially the same error statistics for
absolute ionization energies and chemical shifts. For core excitations,
the absolute energy scale of the spectrum is often irrelevant, and
calculated spectra are simply shifted to match either the first excitation
peak or the ionization edge of the experimental spectrum. This shifting
would be expected to have a similar effect on the error as going from
absolute ionization energies to chemical shifts. One slight difference,
however, is that the shift commonly applied to excitation energies
is edge-specific, while chemical shifts constitute a global rigid
shift for each element.

\subsection{Term Values}

\begin{figure}
\begin{centering}
\thispagestyle{empty}
\begin{tikzpicture}
\begin{axis}[
scale=1.5,
samples=100,smooth,
xmin=-1, xmax=1,
ymin=0, ymax=5,
ytick=\empty,
xlabel={Error w.r.t. CVS-EOM-CCSDT (eV)},
legend style={legend pos=north west}
]
\addplot[draw=red] {gauss(-0.14648701,0.224671857)};
\addlegendentry{CCSD}
\addplot[draw=blue] {gauss(0.172053693,0.089588458)};
\addlegendentry{CCSDT-1}
\addplot[draw=black!50!green] {gauss(0.063232143,0.089132875)};
\addlegendentry{CCSDT-3}
\addplot[draw=orange] {gauss(0.186482143,0.130366207)};
\addlegendentry{CC3}
\addplot[draw=brown] {gauss(-0.005031597,0.065800694)};
\addlegendentry{$\Delta$CCSD}
\addplot[draw=red,dashed] {gauss(0.023632585,0.085325467)};
\addlegendentry{CCSD*}
\addplot[draw=blue,dashed] {gauss(0.169582568,0.37459561)};
\addlegendentry{CCSDR(3)}
\addplot[draw=black!50!green,dashed] {gauss(0.031540642,0.108983698)};
\addlegendentry{CCSD(T)(a)*}
\addplot[draw=black] coordinates {(0,-1) (0,100)};
\end{axis}
\end{tikzpicture}
\par\end{centering}
\caption{\label{fig:term}Normal error distributions for core excitation term
values. Iterative methods are denoted by solid lines, and perturbative
methods by dashed lines. All methods except $\Delta$CCSD are core-valence
separated EOM, e.g. ``CCSD'' = CVS-EOM-CCSD.}
\end{figure}
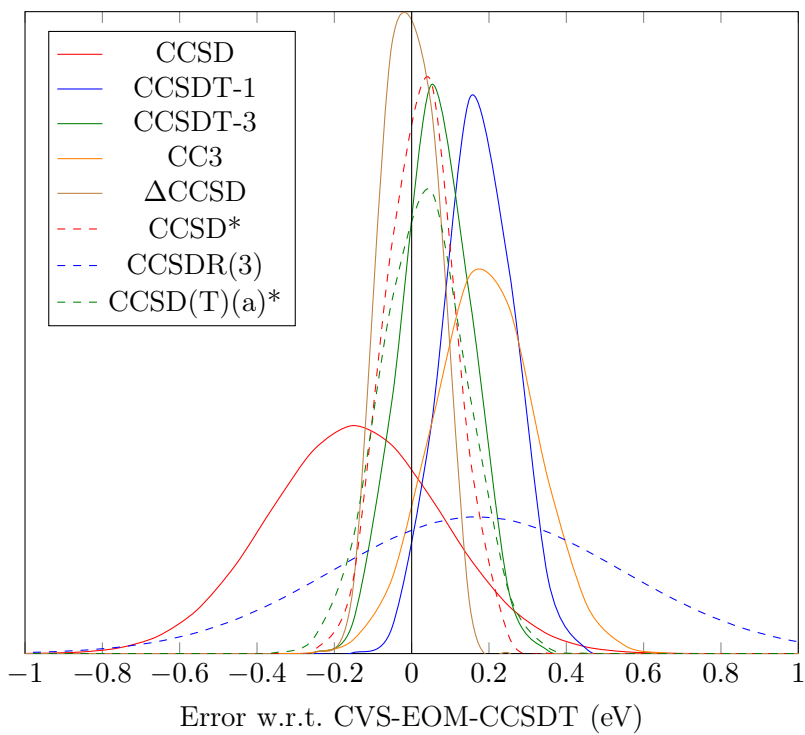

To capture the effect of a system-specific shift on the core excitation
energies, we have recomputed normal error distributions based on the
error in the excitation term value, that is the difference in energy
between the excitation and the corresponding ionization edge. We have
used this measure rather than an experimental shift in order to maintain
an entirely internally-calibrated theoretical benchmark. The term
value normal error distributions are plotted in \figref{term}. For
CCSD, we can see that the mean error is significantly reduced, indicating
that such a shift can indeed reduce the average deviation, but the
standard deviation is essentially unchanged. A similar effect is observed
for CCSD{*}, CCSD(T)(a){*}, and CCSDT-3 (CCSD{*} corr. is exactly
equal to CCSD{*} for term values). For CCSDT-1 and CC3 some reduction
in mean error is observed as before, but the standard deviation is
significantly reduced as well. The most probably cause is that residual
correlation errors (from the triples approximation) in the excited
and ionized states almost precisely cancel, while this error does
not necessarily cancel between different chemical systems. It should
also be noted that the mean errors for these two methods are now the
largest among the tested methods. This indicates a ``consistent inconsistency''
in the correlation treatment of the excited and ionized states, such
that the ionization energy is consistently overestimated compared
to the excitation energies, along with a high degree of error cancellation.
Finally, the CCSDR(3) method shows very odd behavior for term values,
with the standard deviation increasing by almost a factor of four
compared to the vertical excitation energies. As this method also
fared poorly for ionization energies, it seems that unlike CCSDT-1
and CC3 there is no benefit from error cancellation and in fact a
degradation of the error inherited from the poor description of ionization.
As CCSDR(3) and CCSD(T)(a){*} are similar theoretically, it is not
clear what the source of this error is, although a closer look at
the individual errors shows that CCSDR(3) has trouble describing nitrogen
core ionizations in particular.

\section{Conclusions}

Core ionization and core excitation energies for a variety of small
molecules were computed with a selection of both iterative and non-iterative
triple excitation methods, as well as with CVS-EOM-CCSD and $\Delta$CCSD
which neglect triple excitation effects. Comparison to full CVS-EOM-CCSDT
shows that the unsystematic error in CVS-EOM-CCSD can be reduced by
a factor of 2-3 by an approximate inclusion of triples effects, while
systematic error in excitation energies can be largely eliminated
by shifting the spectrum. The $\Delta$CCSD method significantly improves
on CVS-EOM-CCSD by taking orbital relaxation into account explicitly,
and despite a complete lack of triple excitations, is the most accurate
method tested here. Among the approximate triples methods, CVS-EOM-CCSD{*},
optionally augmented by an ionization energy correction from full
CVS-EOM-CCSD(T)(a), is the most accurate method. The iterative CVS-EOM-CCSDT-3
method is also highly accurate, while the similar CVS-EOM-CCSDT-1
and CVS-EOM-CC3 methods are less accurate and in particular seem to
consistently underestimate the gap between the excitation spectrum
and the ionization edge. The CVS-EOM-CCSDR(3) method, while similar
theoretically to other methods tested, does not seem to be suitable
for calculating core excitation and ionization energies.

The ability of approximate triples methods to accurate describe soft
X-ray spectra is an exciting prospect for highly-accurate spectroscopy
in this region, especially as all of the methods tested scale as $\mathscr{O}(n^{6})$
for the excited state calculation. In addition, the CVS-EOM-CCSD{*}
method scales rigorously as $\mathscr{O}(n^{6})$ including the ground
state, and is coincidentally the most accurate EOM method. While the
current study focuses only on energetics, it would also be interesting
to explore approximate methods for calculating transition strengths.
Going beyond energetics also weighs in favor of CVS-EOM-CC approaches
over $\Delta$CCSD, as the derivation of analytic properties in the
latter would be difficult to say the least, in addition to the difficulty
of converging a sufficient number of states with two-determinant coupled
cluster.

\section*{Acknowledgements}

DAM would like to acknowledge a generous start-up package from SMU.

\section*{Disclosure Statement}

No potential conflict of interest was reported by the authors.

\bibliographystyle{tfo}
\bibliography{paper}

\end{document}